\documentclass[pra,aps,twocolumn,floatfix,showpacs,superscriptaddress]{revtex4}
\usepackage{graphicx}
\usepackage{amsmath}
\begin{document}
\title{Tunable superluminal reflection and transmission through a
slab doped with Raman driven atoms}

\author{Yasir Ali}
\email[Correspondence email address: ]{yasiralikhan_25@yahoo.com}
\affiliation{Department of Physics and Applied Mathematics, Pakistan Institute of Engineering and Applied Sciences, Nilore, Islamabad 45650, Pakistan}

\begin{abstract}
A study about the reflection and transmission of an electromagnetic pulse through a slab doped with four-level atomic system has been presented. The doped atoms are considered to be in $N-$configuration with a pump field and a weak probe field. An additional control field is also applied to manipulate the Raman gain of the atomic medium. The propagation of transmitted and reflected pulses at different control field strengths and different slab thickness has been studied. It is found that the transmitted and reflected pulses can be simultaneously subluminal or superluminal. These pulses can be controlled from subluminal to superluminal by adjusting the strength of the control field.
\end{abstract}
\pacs{42.25.Bs, 42.25.Gy, 42.50.Gy}

\maketitle
\section{Introduction}
The study of electromagnetic pulse propagation through a dispersive medium has been a topic of considerable interest during several past decades \cite{R2}. These studies show that, in a dispersive medium light pulse can be slowed down or speeded up {\cite{RYChiao} which is also demonstrated in a series of excellent experiments. For example, the subluminal propagation of light (slow light) is demonstrated in a number of physical systems including atomic vapors \cite{kasapi, schmidt, hau, kash, budker} and solid systems \cite{turukhin, bigelow}. Similarly, there are many interesting demonstrations of superluminal light (pulse with group velocity greater than c or negative)\cite{chu, steinberg, spielmann, wang}. Propagation of optical pulses under anomalous dispersion condition has been studied extensively in absorptive medium \cite{R4, Chu1, Segard, Akulshin} and gain medium \cite{R5, Icsevgi, Picholle, Fisher, Chiao, Chiaob}. 
Steinberg and Chiao discussed dispersionless, highly superluminal propagation without significant gain, loss, distortion or broadening \cite{Steinberg1}. In another study \cite{Arbiv}, it is shown that the group velocity can be controlled from subluminal to superluminal by controlling phases of two applied fields to a $V-$shaped three-level system. Japha et. al. \cite{Japha} discussed the supeluminal propagation through nondissipative media.

Recently, Wang et.al. \cite{Li-Wang} studied the propagation of light pulses through a slab which is doped with two-level and three-level atomic system. The doped atoms are considered to be passive or active. It was shown that the reflected pulses can be controlled from subluminal to superluminal by changing the thickness of the slab or the background dielectric constant of the slab. Moreover, by choosing the type of dopant atom (i.e., absorptive or gain) and with the proper choice of slab thickness, both the reflected and transmitted pulses can be made superluminal. In a related study, Jafari et al. \cite{jafari} considered a slab doped with ladder-type three-level atoms. They have shown that the reflected and transmitted pulses can be controlled from subliminal to superluminal by controlling the intensity of the coupling field or changing the slab thickness. The effect of quantum interference which is present in ladder-type three-level atomic system is also addressed which shows that the reflected and transmitted pulses are phase dependent. 

Recently, there has been much interest in studies based upon stimulated Raman gain systems \cite{R5, sajid}. For example, in an interesting study Li et al. demonstrated using active Raman gain, a fast Kerr phase gate where the probe wave travels superluminally \cite{Li}. In another scheme,  Arkhipkin and Myslivets proposed an interesting method for all-optical transistor \cite{arkh1} using a four-level N-type active Raman gain medium doped in a one-dimensional photonic-crystal. The main advantage is the fact that the attenuation of the prob field can be eliminated due to the Raman gain of the system. As a result, both the subluminal and superluminal propagation can be realized.    

In this paper, the transmission and reflection of a light pulse from a slab doped with a Raman gain medium has been considered. The Raman gain medium essentially consists of four-level N-type atomic system subjected to a pump field, a weak prob field, and a control field. In a recent study, Agarwal and Dasgupta showed that this control field can be used to manipulate the Raman gain process of the four-level system which modifies the optical coherence of the medium \cite{R5}. The gain properties of the system can be obtained from the susceptibility of the atomic medium. In this study, it has been  shown that, the control field is a very effective parameter to manipulate the group velocity of the reflected and transmitted pulses from the doped slab. For example, the reflected and transmitted pulses can be switched from subluminal to superluminal by proper choice of the control field strength. The method of phase time is used to study the transmission and reflection time of pulses through doped slab which is also verified by studying the evolution of the actual envelop of the incident pulse.
\section{Propagation of Light Pulse in a Slab}
A Gaussian light pulse $E_i(z,t)$ incident normally on a nonmagnetic slab of thickness $d$ has been considered. The incident light pulse is traveling in z-direction through the slab which is extended from $z=0$ to $z=d$ whereas the slab is considered to be of infinite length in $xy$ plane. To study the behavior of each frequency component, the incident pulse can be expressed in integral form of its Fourier components $E(z,\omega)$. Upon entering the slab, the incident pulse can be divided into reflected and transmitted parts whose reflection and transmission coefficients can be derived from transfer matrix \cite{R1}
\begin{equation}
{\bf{M}} = \left[ {\begin{array}{*{20}c}
   {\cos (\frac{\omega }{c}n (\omega )\Delta z)} & {\frac{1}{{n (\omega )}}\sin (\frac{\omega }{c}n (\omega )\Delta z)}  \\
   { - n (\omega )\sin (\frac{\omega }{c}n (\omega )\Delta z)} & {\cos (\frac{\omega }{c}n (\omega )\Delta z)}  \\
\end{array}} \right],
\label{E1}
\end{equation}
where $c$ is the velocity of light in vacuum and $n(\omega)$ is frequency dependent refractive index given by $n(\omega)=\sqrt{\varepsilon(\omega)}$ for a nonmagnetic slab. The reflection  and transmission coefficients  i.e., $r(\omega)$ and $t(\omega)$ are given by \cite{R1}

\begin{equation}
r(\omega ) = \frac{{i\left( {n(\omega ) - \frac{1}{{n(\omega )}}} \right)\sin \left( {\frac{{n(\omega )\omega d}}{c}} \right)}}{{2\cos \left( {\frac{{n(\omega )\omega d}}{c}} \right) - \left( {n(\omega ) + \frac{1}{{n(\omega )}}} \right)\sin \left( {\frac{{n(\omega )\omega d}}{c}} \right)}},
\label{E2}
\end{equation}
and
\begin{equation}
t(\omega ) = \frac{2}{{2\cos \left( {\frac{{n(\omega )\omega d}}{c}} \right) - \left( {n(\omega ) + \frac{1}{{n(\omega )}}} \right)\sin \left( {\frac{{n(\omega )\omega d}}{c}} \right)}}.
\label{E3}
\end{equation}
Here it is assumed that outside the slab there is vacuum. Using these reflection and transmission coefficients, reflected and transmitted pulses as $E_r(0,\omega)=r(\omega)E_i(0,\omega)$ and $E_t(d,\omega)=t(\omega)E_i(0,\omega)$ can obtained, respectively. The reflected and transmitted pulses in time domain can be obtained as 
\begin{subequations}
 \begin{align}
 E_r (0,t) & = \int {E_r  (0,\omega )e^{-i\omega t} d\omega }, \label{E4a} \\ 
 E_t (d,t) & = \int {E_t (d,\omega )e^{-i\omega t} d\omega }, \label{E4b}
\end{align}
\label{E4}
\end{subequations} \\
respectively. A narrow incident pulse has been assumed so that distortion in reflected and transmitted pulses is negligibly small. This is a necessary requirement for our peak arriving time calculation. The reflection and transmission coefficients can be written as 
\begin{subequations}
\begin{align}
 r(\omega ) & = |r(\omega )|e^{i\phi _r (\omega )},  \label{E5a}\\ 
 t(\omega ) & = |t(\omega )|e^{i\phi _t (\omega )}.  \label{E5b} 
 \end{align}
\label{E5}
\end{subequations} \\
where $\phi _r (\omega )$ and $\phi _t (\omega )$ are the phases of the reflection and transmission coefficients respectively. The corresponding phase delay times for reflected and transmitted pulse are given by $\tau _r^{peak}  = \frac{{\partial \phi _r (\omega )}}{{\partial \omega }}$ and $\tau _t^{peak}  = \frac{{\partial \phi _t (\omega )}}{{\partial \omega }} $ respectively \cite{R3,R4}. For $\tau _r^{peak}<0$, the pulse reflection  is superluminal whereas the transmitted pulse is superluminal when $\tau _t^{peak}<d/c$. 
\section{Model and Equations}
\label{sec3}
A slab doped with four-level atomic system in $N$ configuration as shown in Fig. \ref{F1} has been considered. The two lower levels $\left| 1 \right\rangle $ and $\left| 3 \right\rangle $ are coupled with the upper levels $\left| 2 \right\rangle $ and $\left| 4 \right\rangle $ via driving fields of frequencies $\nu_1$ and $\nu_c$ which are off-resonant from the atomic transitions by  $\Delta_1$ and $\Delta_c$, respectively. The corresponding strength of these field are given by Rabi frequencies $\Omega_1$ and $\Omega_c$ respectively. A weak probe field of frequency $\nu_p$ is applied to couple lower level $\left| 3 \right\rangle $ with upper level $\left| 2 \right\rangle $ which has Rabi frequency $\Omega_p$ and detuning $\Delta_p$. The Rabi frequency  $\Omega_c$ represents the control field which is used to control Raman gain process in the system.

\begin{figure}[htb]
\includegraphics[scale=0.32]{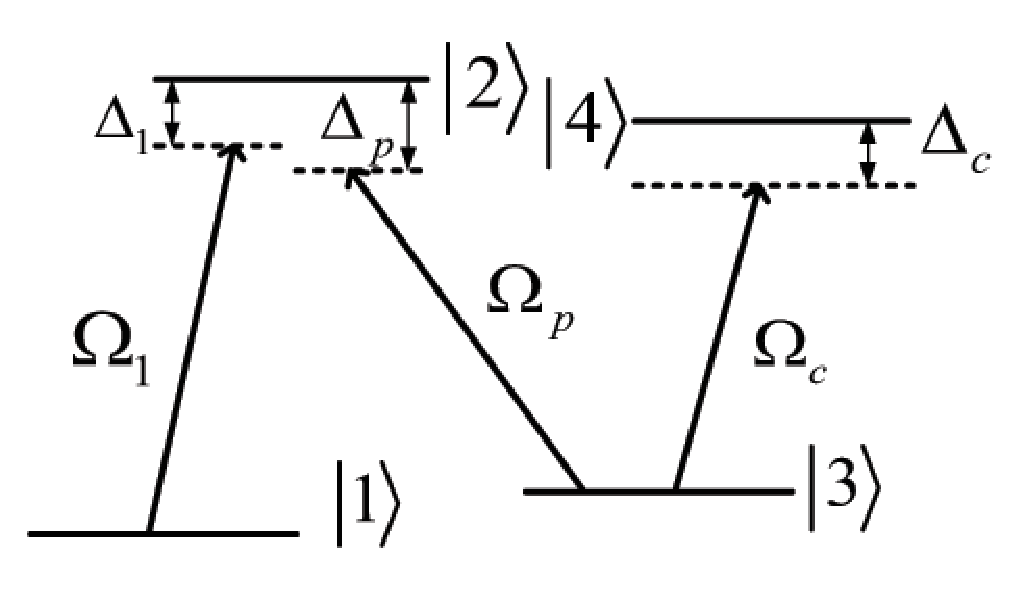}
\caption{ Schematic of a four-level atomic system in $N$ configuration.}
\label{F1}
\end{figure}

Hamiltonian for the system in the dipole and rotating wave approximation is given as 
\begin{equation}
H=H_F+H_I,
\label{E6}
\end{equation}
where free part of Hamiltonian, $H_F$ is given as
\begin{equation}
H_F=\sum {\hbar \omega _i \left| i \right\rangle \left\langle i \right|} ,\;\;\;\;i \in (1,2,3,4),
\label{E7}
\end{equation}
while interaction part of Hamiltonian, $H_I$ is 
\begin{equation}
\begin{array}{l}
 H_I  =  - \frac{\hbar }{2}\left[ {\Omega _1 e^{ - i\nu _1 t} \left| 2 \right\rangle \left\langle 1 \right|} \right. + \Omega _p e^{ - i\nu _p t} \left| 2 \right\rangle \left\langle 3 \right| \\ 
 {\kern 1pt} {\kern 1pt} {\kern 1pt} {\kern 1pt} {\kern 1pt} {\kern 1pt} {\kern 1pt} {\kern 1pt} {\kern 1pt} {\kern 1pt} {\kern 1pt} {\kern 1pt} {\kern 1pt} {\kern 1pt} {\kern 1pt} {\kern 1pt} {\kern 1pt} {\kern 1pt} {\kern 1pt} {\kern 1pt} {\kern 1pt} {\kern 1pt} {\kern 1pt} {\kern 1pt} {\kern 1pt} {\kern 1pt} {\kern 1pt} {\kern 1pt} {\kern 1pt} {\kern 1pt} {\kern 1pt} {\kern 1pt} {\kern 1pt} \left. { + \Omega _c e^{i\nu _c t} \left| 4 \right\rangle \left\langle 3 \right| + H.c.} \right]. \\ 
 \end{array}
 \label{E8}
\end{equation}
Here in this study, the effects of the Raman gain, medium doped in a slab, on the transmission and reflection of incident electromagnetic pulse is point of interest. This can be obtained from the susceptibility, which is given by \cite{R5}

\begin{equation}
\chi  = \frac{{2N\left| {\vec d_{23} } \right|^2 }}{{\hbar \varepsilon_{0} \Omega _p }}\rho _{23} ,
\label{E10}
\end{equation}
where $N$ is the number density of the atomic medium used in doping of slab and $\left| {\vec d_{23} } \right|$ is the magnitude of the dipole matrix element between levels $\left| 2 \right\rangle$ and $\left| 3 \right\rangle$.
By using density matrix approach while keeping terms of all orders in the control field, the nonlinear susceptibility is found to be dependent on control field \cite{R5}
\begin{equation}
\chi  = \beta \frac{\Omega _1^2}{8} D,
\label{E11}
\end{equation}
where $\beta  = \frac{{2N\left| {\vec d_{23} } \right|^2 }}{{\hbar \varepsilon_{0}}}$. The parameter $D$ is given by 
\begin{equation}
\begin{array}{l}
 D = \frac{{ - i}}{A}\left[ {\frac{{2\Gamma _{21} \left\{ {\Gamma _{24}  - i\left( {\Delta _p  - \Delta _c } \right)} \right\}}}{{\left( {\gamma _{12}  + \gamma _{32} } \right)\left( {\Gamma _{12}^2  + \Delta _1^2 } \right)}}} \right. +  \\ 
 \left. {\frac{{\left\{ {\Gamma _{24}  - i\left( {\Delta _p  - \Delta _c } \right)} \right\}\left\{ {\Gamma _{41}  - i\left( {\Delta _p  - \Delta _1  - \Delta _c } \right)} \right\} - {\raise0.7ex\hbox{${\left| {\Omega _c } \right|^2 }$} \!\mathord{\left/
 {\vphantom {{\left| {\Omega _c } \right|^2 } 4}}\right.\kern-\nulldelimiterspace}
\!\lower0.7ex\hbox{$4$}}}}{{\left( {\Gamma _{21}  + i\Delta _1 } \right)\left[ {\left\{ {\Gamma _{13}  - i\left( {\Delta _p  - \Delta _1 } \right)} \right\}\left\{ {\Gamma _{41}  - i\left( {\Delta _p  - \Delta _1  - \Delta _c } \right)} \right\} + {\raise0.7ex\hbox{${\left| {\Omega _c } \right|^2 }$} \!\mathord{\left/
 {\vphantom {{\left| {\Omega _c } \right|^2 } 4}}\right.\kern-\nulldelimiterspace}
\!\lower0.7ex\hbox{$4$}}} \right]}}} \right], \\ 
 \end{array}
 \label{E12}
\end{equation}
and
\begin{equation}
A = \left( {\Gamma _{23}  - i\Delta _p } \right)\left\{ {\Gamma _{24}  - i\left( {\Delta _p  - \Delta _c } \right)} \right\} + {\raise0.7ex\hbox{${\left| {\Omega _c } \right|^2 }$} \!\mathord{\left/
 {\vphantom {{\left| {\Omega _c } \right|^2 } 4}}\right.\kern-\nulldelimiterspace}
\!\lower0.7ex\hbox{$4$}},
\label{E13}
\end{equation}
where $\Gamma_{ij}$ is dephasing parameter and $\gamma_{ij}$ are spontaneous decay rates from $j \;\;(j \in 1,3)$ to $i \;\;(i \in 2,3)$. Here $\rho _{23}=\Omega_1^2 \Omega_p D/8$ is the optical coherence modified by the control field. It is shown by Agarwal and Dasgupta \cite{R5} that the Raman gain process can be manipulated coherently by the control field $\Omega_c$ such that normal dispersion regime can be switched to the anomalous dispersion. Interest of the study is to see the effect of this controllable parameter when light pulse passes through a glass slab doped with such atomic system.

\begin{figure}[ht]
\begin{center}
\includegraphics[width=3.450in]{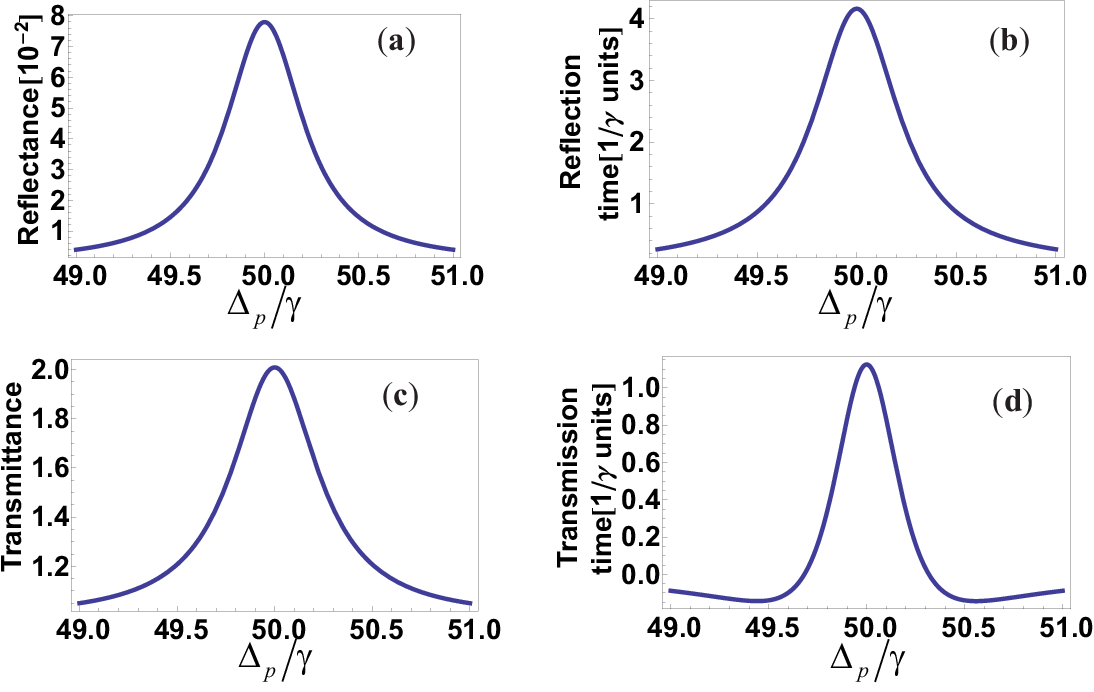}
\end{center}
 \caption{ Reflectance (a) and transmittance (c) vs $\Delta_p/\gamma$ along with corresponding phase times (b) \& (d), respectively for $\Omega_c=1.5\gamma$. Here $d=2m\lambda_0/(4\sqrt{\varepsilon_b})$ and $m=1.5\times10^3$ while other parameters are the same as defined in the text.}
  \label{F2}
 \end{figure}

\begin{figure}[b]
\begin{center}
\includegraphics[width=3.450in]{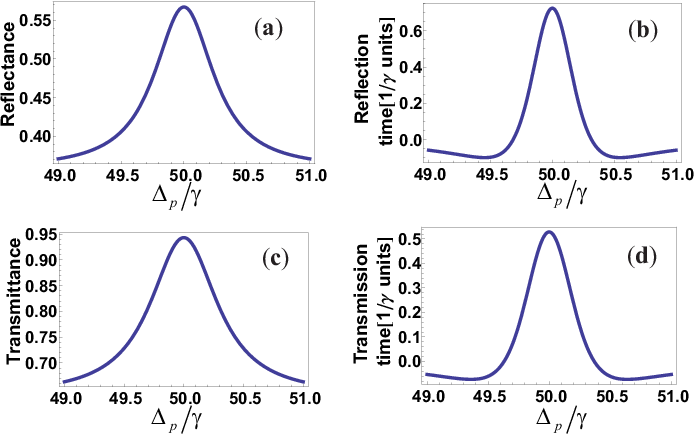}
\end{center}
 \caption{Reflectance (a) and transmittance (c) vs $\Delta_p/\gamma$ along with corresponding phase times (b) \& (d), respectively for $\Omega_c=1.5\gamma$. Here $d=(2m+1)\lambda_0/(4\sqrt{\varepsilon_b})$ and $m=1.5\times10^3$ while other parameters are the same as defined in the text.}
 \label{F3}
 \end{figure}

\section{Results and Discussions}
An incident pulse having Gaussian distribution in time which is normally incident on slab at $z=0$ has been considered. Electric field of this Gaussian pulse can be written as 
\begin{equation}
E_i (0,t) = A_0 \exp \left[ { - \frac{{t^2 }}{{2t_0^2 }}} \right]e^{ - i\omega _0 t},
\label{E14}
\end{equation}
with its Fourier transform given as
\begin{equation}
E_i (0,\omega ) = \frac{{A_0 \tau _0 }}{{2\sqrt \pi  }}\exp \left[ { - \frac{{t_0^2 (\omega  - \omega _0 )^2 }}{2}} \right],
\label{E15}
\end{equation}
where $t_0$ is temporal width, $A_0$ is amplitude of Gaussian pulse and $\omega_0$ is carrier frequency of the pulse. Here $t_0=20\times 10^{-6}$ sec and $\omega_0 \approx 10^{15}$Hz are taken \cite{Li-Wang}.

The incident pulse is divided into reflected and transmitted pulses upon entering the slab. Reflection and transmission coefficients of slab are given by Eqs. (\ref{E2}) and (\ref{E3}). The refractive index for the doped slab can be written as $n(\omega)=\sqrt{\varepsilon_b+\chi}$ where the dielectric susceptibility $\chi$ depends on the dopant (i.e., four-level atoms) (See Eq. (\ref{E11})) and the background dielectric constant is chosen to be $\varepsilon_b=4.0$.

Here, the results of our numerical simulations for the transmittance (reflectance) and transmission time (reflection time) has been presented. The detuning parameters as $\Delta_c=0$ and $\Delta_1=50\gamma$ has been chosen. The other parameters are $\Gamma_{43}=\Gamma_{41}=\Gamma_{23}=\Gamma_{21}=2.01\gamma, \Gamma_{24}=4.01\gamma, \Gamma_{13}=0.01\gamma,\gamma_{12}=\gamma_{32}=\gamma_{34}=\gamma_{14}=2\gamma$  where $\gamma\sim 10^6/sec$. The value of parameter $\beta$ is chosen to be $0.16\gamma$. These parameters remain fixed during the rest of the analysis. First the smaller strength of Rabi frequency for control field i.e., $\Omega_c=1.5\gamma$ has been considered. In Figs. \ref{F2}(a) and \ref{F2}(c) reflectance and transmittance are shown for slab thickness $d=2m\lambda_0/(4\sqrt{\varepsilon_b})$. In this case both the reflectance and transmittance have peaks at $\Delta_p=50\gamma$. In Figs. \ref{F2} (b) and (d), the plots of reflection and transmission times has been shown. The values of the transmission and reflection time can be obtained from the peak position of the transmitted and reflected pulse which are given by $\tau_r=4.16173\:/ \:\gamma$ and $\tau_t=1.12556\:/ \:\gamma$ respectively. These values are higher than the corresponding propagation time of a pulse in vacuum therefore, the reflection and transmission in this case is subluminal. Changing the length of slab to $d=(2m+1)\lambda_0/(4\sqrt{\varepsilon_b})$ (i.e., off-resonance case) only changes numerical values of reflection and transmission times i.e., $\tau_r=0.724678\:/\:\gamma$ and $\tau_t=0.528631\:/\:\gamma$, and does not change subluminality of reflection and transmission as evident in Fig. \ref{F3}.

\begin{figure}[htb]
\begin{center}
\includegraphics[width=3.450in]{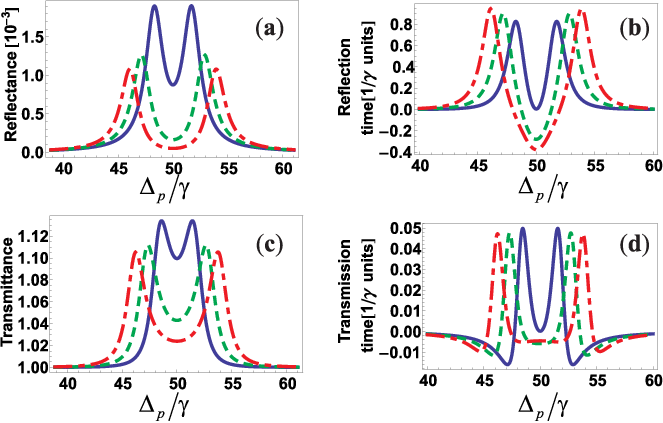}
\end{center}
\caption{ Reflectance (a) and transmittance (c) vs $\Delta_p/\gamma$ along with corresponding phase times (b) \& (d), respectively for three different choices of $\Omega_c$. Here solid line corresponds to $\Omega_c=4\gamma$, dashed line corresponds to $\Omega_c=6\gamma$ and dotted-dashed line corresponds to $\Omega_c=8\gamma$ for $d=2m\lambda_0/(4\sqrt{\varepsilon_b})$ and $m=1.5\times10^{3}$. The rest of the parameters are the same as defined in the text}
\label{F4}
\end{figure}

Next, the effects of higher strength of control field Rabi frequency $\Omega_c$ on reflection and transmission has been studied as shown in Figs. \ref{F4} and \ref{F5}. In Fig. \ref{F4}, reflectance (transmittance) and reflection time (transmission time) are plotted  using slab thickness $d=2m\lambda_0/(4\sqrt{\varepsilon_b})$ for different strengths of the control field. In  Figs. \ref{F4}(a) and \ref{F4}(c), the solid line corresponds to $\Omega_c=4.0\gamma$, where two gain peaks and one dip is present in reflectance and transmittance. Despite of this dip, reflection and transmission are subluminal. This is evident from the corresponding phase times given by $\tau_r=5.210\times10^{-3}\:/\:\gamma$ (which is positive) and $\tau_t=2.383\times10^{-4}\:/\:\gamma$ (which is greater than $d/c$) as shown by solid curve in Figs. \ref{F4}(b) and \ref{F4}(d). By increasing the strength of control field to $\Omega_c=6\gamma$ (dashed line in Fig. \ref{F4}), the dips of reflectance and transmittance becomes more deeper than previous case. Similarly reflection and transmission times also have deeper dips now. The minimum values of these times (the center of dip at $\Delta_c=50\gamma$) are negative and given by $\tau_r=-0.279\:/\:\gamma$ and $\tau_t=-5.779\times10^{-3}\:/\:\gamma$ as shown in Figs. \ref{F4}(b) and \ref{F4}(d). This negative phase time corresponds to superluminality. It must be noted that the transition from subluminal regime to superluminal regime is achieved by only increasing the strength of control field. This is the most important result of this work. Further increase in control field does not affect the superluminality but only changes numerical values of reflection and transmission times. For instance, when $\Omega_c=8\gamma$ (dotted-dashed line in Fig. \ref{F4}) then these times are $\tau_r= -0.375\:/\:\gamma$ and $\tau_t=-4.403\times10^{-3}\:/\:\gamma$, respectively.

Next, the thickness of the slab has been changed to $d=(2m+1)\lambda_0/(4\sqrt{\varepsilon_b})$ and the effect of increasing control field on reflected and transmitted pulses has been studied. For $\Omega_c=4\gamma$ (solid curve in Fig. \ref{F5}), reflectance and transmittance have dips at $\Delta_p=50\gamma$. These dips correspond to minimum phase time at center $\Delta_p=50\gamma$ of reflected and transmitted pulses, which has positive values ($\tau_r=1.543\times10^{-4}\:/\:\gamma$ and $\tau_t=1.473\times10^{-4}\:/\:\gamma$). Therefore, reflection and transmission are subluminal as obtained in the earlier case. Further increase in the strength of the control field produces similar effects as discussed above with reference to Fig. \ref{F4}. It is found that the superluminality does not change by changing the length of slab. It is only the magnitude of reflectance and transmittance that changes. Similarly, reflection and transmission times remain negative for other choices of $\Omega_c$ as shown in Fig. \ref{F5}. These times are $\tau_r=-3.720\times10^{-3}\:/\:\gamma$ and $\tau_t=-3.643\times10^{-3}\:/\:\gamma$ for $\Omega_c=6\gamma$ and $\tau_r=-2.828\times10^{-3}\:/\:\gamma$ and $\tau_t=-2.794\times10^{-3}\:/\:\gamma$ for $\Omega_c=8\gamma$, respectively.

\begin{figure}[ht]
\begin{center}
\includegraphics[width=3.450in]{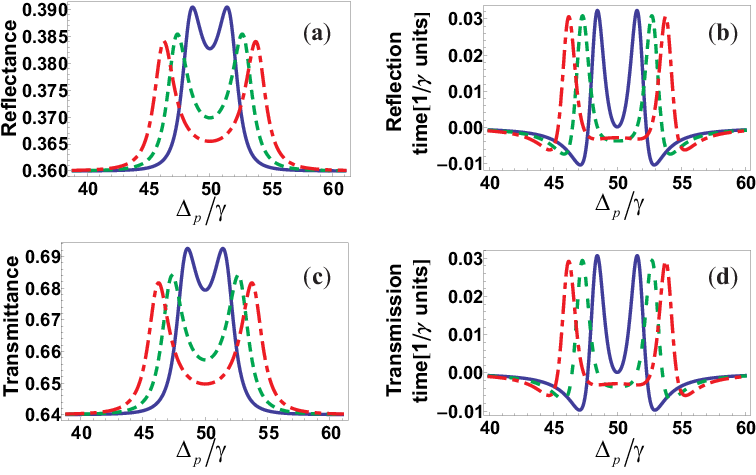}
\end{center}
\caption{ Reflectance (a) and transmittance (c) vs $\Delta_p/\gamma$ along with corresponding phase times (b) \& (d), respectively for three different values of $\Omega_c$. Here solid line corresponds to $\Omega_c=4\gamma$, dashed line corresponds to $\Omega_c=6\gamma$ and dotted-dashed line corresponds to $\Omega_c=8\gamma$ for $d=(2m+1)\lambda_0/(4\sqrt{\varepsilon_b})$ and $m=1.5\times10^{3}$. The rest of the parameters are the same as defined in the text.}
\label{F5}
\end{figure}

\begin{figure}[ht]
\begin{center}
\includegraphics[width=3.450in]{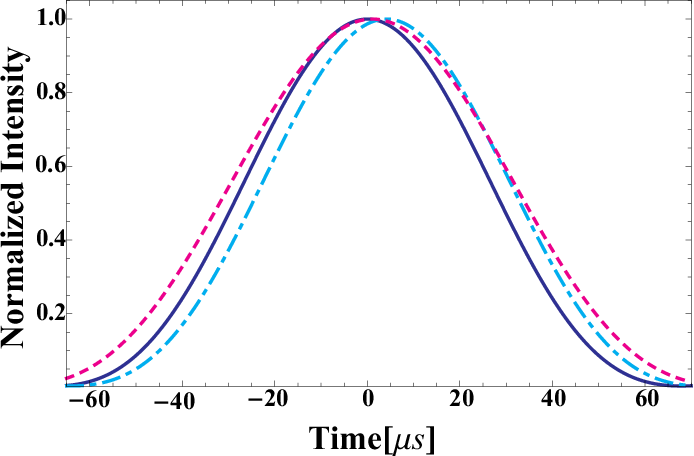}
\end{center}
\caption{ Normalized intensity of the reflected and transmitted pulses vs time for $\Omega_c=1.5\gamma$ while length of the slab is taken as $d=2m\lambda_0/(4\sqrt{\varepsilon_b})$. Solid line corresponds to reference pulse, dotted-dashed line corresponds to reflected pulse and dashed line corresponds to transmitted pulse.}
\label{F6}
\end{figure}

\begin{figure}
\includegraphics[width=3.450in]{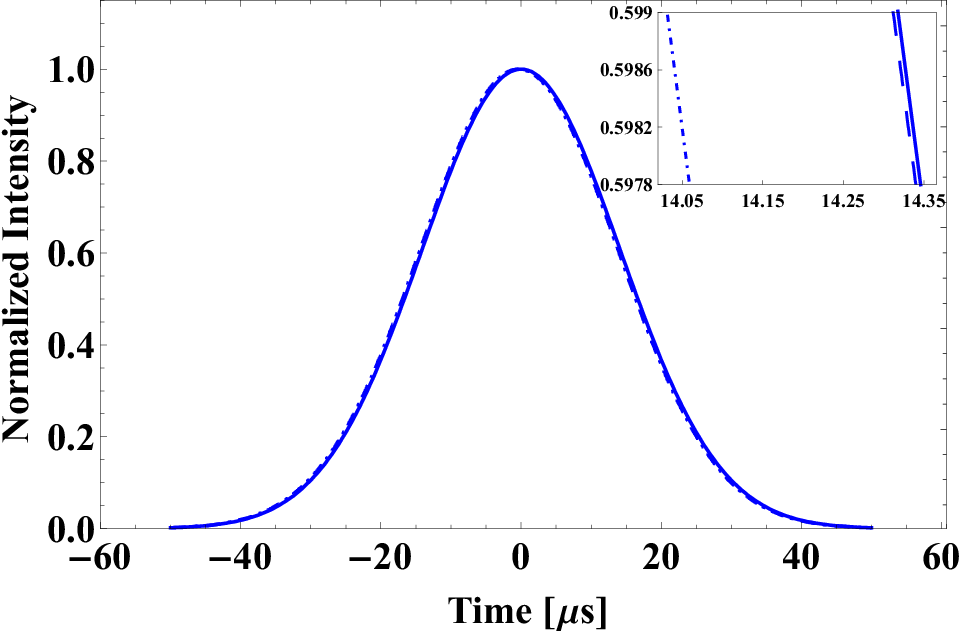}
 \caption{Normalized intensity of the reflected and transmitted pulses vs time for $\Omega_c=6\gamma$ while length of the slab is taken as $d=2m\lambda_0/(4\sqrt{\varepsilon_b})$. Solid line corresponds to reference pulse, dotted-dashed line corresponds to reflected pulse and dashed line corresponds to transmitted pulse.}
 \label{F7}
\end{figure}
In order to verify the above results, integration of Eqs. (\ref{E4a}) \& (4b) can be performed using $t_0=20\mu s$ and $d=2m\lambda_0/(4\sqrt{\varepsilon_b})$ and keeping other parameters same as defined in section-\ref{sec3}. In Fig. \ref{F6} the normalized intensity of the reflected and transmitted pulses is plotted as a function of time for $\Omega_c=1.5\gamma$ along with the actual pulse as a reference pulse. It is clear that both reflected (dotted-dashed line) and transmitted (dashed line) pulses are subluminal. Peaks of these pulses appear after the peak of reference (solid line) pulse. Peak position obtained for reflected and transmitted pulses are at $\tau_r=4.1\:\:/\gamma$ and $\tau_t=1.1\:\:/\gamma$, respectively. It can be seen that these values of reflected and transmitted time obtained by integration of actual envelope are in good agreement with those obtained using phase time definition (see Fig. \ref{F2}(b) and (d)). Same has also been shown this for $\Omega_c=6\gamma$ in Fig. \ref{F7}. In this case reflected (dotted-dashed line) and transmitted (dashed line in Fig. \ref{F7}) pulses are superluminal and their peaks appear before the peak of reference pulse (solid line in Fig. \ref{F7}). The corresponding peak times for these pulses are negative and given by $\tau_r=-0.28\:\:/\gamma$ and $\tau_t= -6\times10^{-3}\:\:/\gamma$, respectively. Since the difference in peak times for pulses is very small to be observed therefore a zoomed portion of this has been shown in the inset of Fig. \ref{F7}. A comparison of numerical values of reflection and transmission times obtained here shows a good agreement with the results presented in Fig. \ref{F4}(b) and (d) (dashed curve). In this way, the results obtained through the phase time definition are validated with those obtained through exact numerical simulation of wave packet propagation.
\section{Conclusions}
In this paper, a scheme for the simultaneous superluminal propagation of reflected and transmitted electromagnetic pulses through a slab doped with four-level atomic system has been presented and discussed. The doped atoms are considered to be four-level system in N configuration with pump and weak prob fields. In addition, a coherent control field is also applied to manipulate the Raman gain process. As a result the transition between normal and anomalous dispersion regimes can be controlled efficiently. The incident electromagnetic pulse is assumed to be of Gaussian shape. The reflection and transmission properties of the incident Gaussian pulse are studied. It is found that the phase time for both the reflected and transmitted pulses exhibits both the subluminal and superluminal behavior.  By plotting the reflectance (transmittance) and the corresponding reflection (transmission) phase time for various choices of control field, it is shown that the reflection and transmission of the pulse can be tuned from subluminal to superluminal by controlling the strength of the control field alone. These results are then verified by the numerical simulation of the pulse propagation.
\section{Acknowledgements}
It is a pleasure to thank Dr. Shahid Qamar (PIEAS, Islamabad), Dr. Sajid Qamar (COMSATS, Islamabad) and Dr. Muhammad Irfan (PIEAS, Islamabad) for their guidance, helpful comments and corrections at multiple times during this work.


\begin{thebibliography}{}
 
\bibitem{R2} Brillouin L 1960 Wave Propagation and Group Velocity (Academic, New York)
\bibitem{RYChiao} Chiao R Y, Kozhekin A E and Kurizki G 1996 Phys. Rev. Lett. {\bf 77} 1254
\bibitem{kasapi} Kasapi A, Jain M, Yin G Y and Harris S E 1995 Phys. Rev. Lett. {\bf 74} 2447
\bibitem{schmidt} Schmidt O, Wynands R, Hussein Z and Meschede D 1996 Phys. Rev. A {\bf 53} R27
\bibitem{hau} Hau L V, Harris S E, Dutton Z and Behroozi C H 1999 Nature (London)  {\bf 397} 594
\bibitem{kash} Kash M M, Sautenkov V A, Zibrov A S, Hollberg L, Welch G R, Lukin M D, Rostovtsev Y, Fry E S and Scully M O 1999 Phys. Rev. Lett. {\bf 82} 5229
\bibitem{budker} Budker D, Kimball D F, Rochester S M and Yashchuk V V 1999 Phys. Rev. Lett. {\bf 83} 1767
\bibitem{turukhin} Turukhin A V, Sudarshanam V S, Shahriar M S, Musser J A, Ham B S and Hemmer P R 2001 Phys. Rev. Lett. {\bf 88} 023602
\bibitem{bigelow} Bigelow M S, Lepeshkin N N and Boyd R W 2003 Phys. Rev. Lett. {\bf 90} 113903
\bibitem{chu} Chu S and Wong S 1982 Phys. Rev. Lett. {\bf 48} 738
\bibitem{steinberg} Steinberg A M, Kwiat P G and Chiao R Y 1993 Phys. Rev. Lett. {\bf 71} 708
\bibitem{spielmann} Spielmann Ch, Szipocs R, Stingl A and Krausz F 1994 Phys. Rev. Lett. {\bf 73} 2308 
\bibitem{wang} Wang L J, Kuzmich A and Dogariu A 2000 Nature (London) {\bf 406} 277
\bibitem{R4} Garrett C G B and McCumber D E 1970 Phys. Rev. A {\bf 1} 305
\bibitem{Chu1} Chu S and Wong S 1982 Phys. Rev. Lett. {\bf 48} 738
\bibitem{Segard} Segard B and Macke B 1985 Phys. Lett. {\bf 109A} 213
\bibitem{Akulshin} Akulshin A M, Barreiro S and Lezama A 1999 Phys. Rev. Lett. {\bf 83} 4277
\bibitem{R5} Agarwal G S and Dasgupta S 2004 Phys. Rev. A {\bf 70} 023802
\bibitem{Icsevgi} Icsevgi A and Lamb W E 1969 Phys. Rev. {\bf 185} 517
\bibitem{Picholle} Picholle E, Montes C, Leycuras C, Legrand O and Botineau J 1991 Phys. Rev. Lett. {\bf 66} 1454
\bibitem{Fisher} Fisher D L and Tajima T 1993 Phys. Rev. Lett. {\bf 71} 4338
\bibitem{Chiao} Chiao R Y 1993 Phys. Rev. A {\bf 48} R34
\bibitem{Chiaob} Chiao R Y and Steinberg A M 1997 in Progress in Optics, edited
by E. Wolf (Elsevier, Amsterdam) p. 345.
\bibitem{Steinberg1} Steinberg A M and Chiao R Y 1994 Phys. Rev. A {\bf 49} 2071
\bibitem{Arbiv} Bortman-Arbiv D, Wilson-Gordon A D and Friedmann H 2001  Phys. Rev. A {\bf 63} 043818
\bibitem{Japha} Japha Y and Kurizki G 1996 Phys. Rev. A {\bf 53} 586
\bibitem{Li-Wang} Wang L G, Chen H and Zhu S Y 2004 Phys. Rev. E {\bf 70} 066602
\bibitem{jafari} D. Jafari, M. Sahrai, H. Motavalli, and M. Mahmoudi, Phys. Rev. A {\bf 84}, 063811 (2011).
\bibitem{sajid} S. Qamar, A. Mehmood, and S. Qamar,  Phys. Rev. A {\bf 79}, 033848 (2009).
\bibitem{Li} R. B. Li, L. Deng, and E.W. Hagley, Phys. Rev. Lett. {\bf 110}, 113902 (2013).
\bibitem{arkh1} V. G. Arkhipkin and S. A. Myslivets, Phys. Rev. A {\bf 88}, 033847 (2013).
\bibitem{R1} N.-H. Liu, S.-Y. Zhu, H. Chen and X. Wu, Phys. Rev. E, {\bf 65}, 046607 (2002).
\bibitem{R3} L. G. Wang, N. H. Liu, Q. Lin, and S. Y. Zhu, Phys. Rev. E {\bf 68}, 066606 (2003).



\end{thebibliography}
\end{document}